\documentclass[10pt, conference, letterpaper]{IEEEtran}
\usepackage{amsmath,amsthm,dsfont}
\usepackage{tabularx,multirow}
\usepackage{graphicx,subfigure,comment}
\usepackage[table]{xcolor}
\usepackage{url} 
\usepackage{dsfont}
\usepackage{cite}
\usepackage{siunitx}
\usepackage{physics} 
\usepackage{pstool}
\usepackage{enumitem}
\usepackage{algpseudocode}
\usepackage[Algorithm]{algorithm}
\usepackage[draft]{hyperref}
\usepackage{autonum} 
\usepackage{mathtools}

\setlist[itemize]{leftmargin=*}
\setlist[enumerate]{leftmargin=*}

\newcommand{\vRoute}{{OKpi}}
\newcommand{\Fig}[1]{Fig.~\ref{fig:#1}}

\newcommand{\Lem}[1]{Lemma~\ref{lem:#1}}

\newcommand{\Sec}[1]{Sec.~\ref{sec:#1}}
\newcommand{\Tab}[1]{Tab.~\ref{tab:#1}}
\newcommand{\Eq}[1]{(\ref{eq:#1})}

\newcommand{\ind}[1]{\mathds{1}_{#1}}

\newcommand{\Ac}{\mathcal{A}}

\newcommand{\Cc}{\mathcal{C}}
\newcommand{\Kc}{\mathcal{K}}

\newcommand{\Ic}{\mathcal{I}}
\newcommand{\Lc}{\mathcal{L}}
\newcommand{\Tc}{\mathcal{T}}
\newcommand{\Vc}{\mathcal{V}}
\newcommand{\Wc}{\mathcal{W}}
\newcommand{\Ec}{\mathcal{E}}

\newcommand{\Sc}{\mathcal{S}}

\newtheorem{property}{Property}

\newtheorem{lemma}{Lemma}

\allowdisplaybreaks

\begin{document}

\title{
{\vRoute}: All-KPI Network Slicing \\
Through Efficient Resource Allocation 
} 

\author{J. Mart\'in Per\'ez$^\dagger$, F. Malandrino$^\ddagger$, C. F. Chiasserini$^{\ast,\ddagger}$, C. J. Bernardos$^\dagger$\\
$\dagger$ Universidad Carlos III de Madrid, Spain -- $\ddagger$ IEIIT-CNR, Torino, Italy  -- 
$\ast$ Politecnico di Torino, Torino, Italy}

\maketitle

\begin{abstract}
Networks can  now process data as well as transporting it; it follows that they can  support multiple services, each requiring different key performance indicators (KPIs). Because of the former, it is critical to efficiently allocate network and computing resources to provide the required services, and, because of  the latter, such decisions must jointly consider all KPIs targeted by a service. Accounting for newly introduced KPIs (e.g., availability and reliability) requires tailored  models and solution strategies, and has been  conspicuously neglected by existing works, which are instead built around traditional metrics like throughput and latency. We fill this gap by presenting a novel methodology and  resource allocation scheme, named {\vRoute}, which enables high-quality selection of radio points of access as well as VNF (Virtual Network Function) placement and data routing, with polynomial computational complexity. {\vRoute} accounts for {\em all} relevant KPIs required by each service, and for any available resource from the fog to the cloud. We prove several important properties of  {\vRoute} and evaluate its performance in two real-world scenarios, finding it to closely match the optimum.
\end{abstract}

\section{Introduction}
\label{sec:intro}

Network Function Virtualization (NFV) enables mobile networks to expand their capabilities beyond data transport and to support  software-based applications on demand. 
Under such a paradigm, third party industries (``verticals'') specify their services  through a graph of virtual network functions (VNFs), then it is the mobile network's task to run such services.  
This requires {\em selecting}  suitable radio points of access (PoAs) as well as {\em placing} and {\em connecting} the VNFs across computing and network resources\footnote{Although memory and storage resources have been omitted for brevity, our framework can handle them as well.}, in order to deliver the vertical services with the required  quality of service. Importantly, different resources can be employed to achieve this goal, ranging from those in the cloud to the ones at the edge of the network infrastructure (i.e., through multi-access edge computing (MEC)) or in the fog (i.e., in  devices such as smartphones, vehicles, robots).  
Note that {\em which} and {\em how many} resources are used is a critical issue, as their associated cost, performance, and availability vary significantly. This is confirmed by \cite{etsi-mec-wp2,zhao2018benders,malandrino2019getting,sang2017provably}, highlighting how cost is a grave  concern for   both the mobile operators owning the cellular infrastructure  and the verticals paying for the services mobile systems should support. 

\begin{figure}
\centering
\includegraphics[width=1\columnwidth]{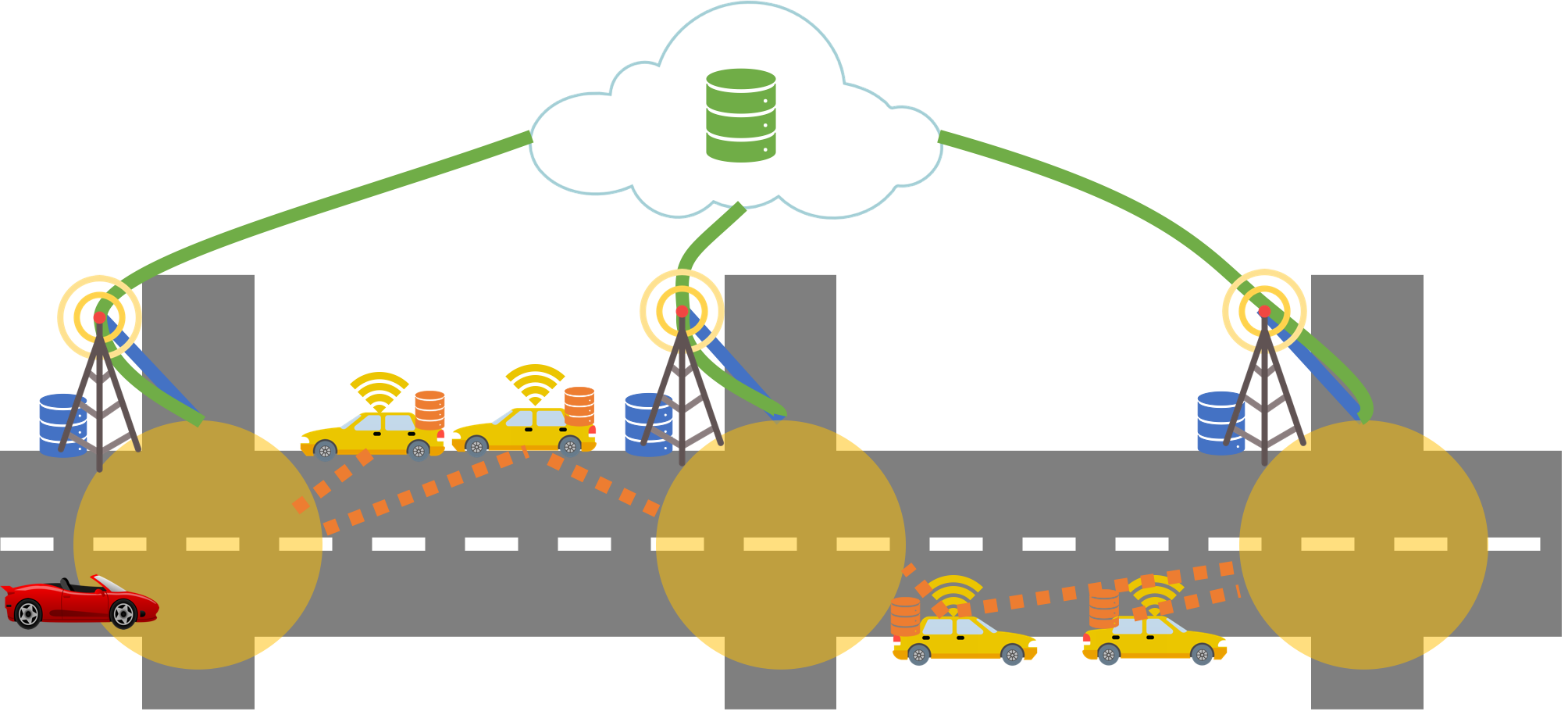}
\caption{\label{fig:fresco} As per geographical availability requirements, a safety service (e.g., warning vehicles on a collision course~\cite{ica_CAMBIAPERCAMERAREADY}) must be provided at the highlighted intersections with high reliability. This can be obtained by deploying: (a) four service instances at parked cars (fog resources), at lower cost but also lower reliability, hence, needing redundancy to meet reliability constraints (orange option); (b) three  instances at the points of access (PoA), e.g., eNBs, covering the intersections (MEC resources, blue option);  (c) deploying only one instance in the cloud, but with larger delay (green option). 
} 
\end{figure}

In spite of the fact that the VNF placement problem
has been already addressed in the  literature (see \Sec{relwork} for a more in-depth discussion), virtually all existing approaches only focus on throughput and  latency as  performance metrics, ignoring what is one of the most disruptive innovations of 5G. Indeed,  
new generation networks have been conceived with the goal of serving multiple use cases (summarized in the ubiquitous ITU ``pyramid'' \cite{pyramid}) whose requirements are both diverse and heterogeneous. {\em Diverse} reflects the fact that, for example, the latency requirements of different use cases can vary by several orders of magnitude.  {\em Heterogeneous} refers to the important fact that 5G introduces several new performance metrics (or KPIs, Key Performance Indicators), including service availability (in both space and time) and service reliability, as exemplified in \Fig{fresco}.
These new KPIs are all but ignored by existing VNF placement algorithms, which may thus be unable to honor the verticals' requirements. We  underline that, as discussed in the next section, accounting for all relevant KPIs requires introducing a new problem formulation and a new solution, which cannot be a mere extension of previous work. 

Furthermore, existing studies have limited or no support for several specific aspects of slicing-based networks, including (i) the opportunity to reuse existing VNF instances (\hspace{-0.1mm}\cite{malandrino2019getting} reuses VNFs within a single data center, while  \cite{lukovszki2018approximate} only focuses on cost), (ii) the possibility of combining cloud-based and MEC-based services (with the exception of \cite{cohen2019access}, which however only deals with caching), and (iii) the need to make decisions on how to place {\em and} connect VNFs, thus jointly addressing VNF placement and data routing (\hspace{-0.1mm}\cite{draxler2018jasper,poularakis2019joint,xu2018joint,agarwal2019vnf} do so, but without considering PoAs or VNF re-usage, and  under some limiting assumptions, e.g., on the number of VNF instances).

{\bf Our contribution and methodology.}  We fill this gap by introducing {\vRoute}, an efficient framework  able to 
create high-quality, end-to-end network slices. 
Specifically, {\vRoute} advances the state-of-the-art in the following main ways:
\begin{description}
    \item[{\em (i)}] it effectively tackles the 5G-defined network slicing KPIs;  
    \item[{\em (ii)}] it leverages  fog, MEC, and cloud resources, allowing VNFs to be placed at any layer of the network topology;   
   \item[{\em (iii)}] it accounts for the fact that already-deployed VNF instances can be reused for newly-requested services\footnote{This is possible if the services share a common subset of VNFs and no service isolation constraints exist.};   
    \item[{\em (iv)}] in such a general setting, it makes {\em joint} decisions on PoA selection, VNF placement, and traffic routing, which  minimize the cost of the resources, thus addressing both mobile operators and verticals' concerns;   
\item[{\em (v)}] it exhibits a low, namely, polynomial, complexity. 
\end{description} 
We remark that, not only the problem we pose is novel, but also our methodology to solve it blends together graph theory and optimization, in a unique fashion. In particular, we describe possible decisions through a graph that reflects their effect on KPIs. Such a graph is then translated into a multi-dimensional expanded graph, which allows us to efficiently find feasible decisions leveraging simple shortest-path algorithms.  The expanded graph can be built with different levels of detail and size, which results into a tuneable tradeoff between computational complexity and optimality.

In the remainder of the paper,  we first review related works in \Sec{relwork},  highlighting which KPIs they consider and the novelty of our study. After that, we introduce the system model in \Sec{model}, and the problem formulation in \Sec{problem}. The {\vRoute} solution and algorithm are described in \Sec{algo};  then \Sec{analysis} proves several relevant properties of {\vRoute} and discusses its computational complexity. \Sec{results} compares {\vRoute} against the optimum in a small-scale, yet practically relevant robotics scenario, and shows its performance in a real-world, large-scale automotive scenario. Finally, \Sec{conclusion} concludes the paper and highlights current research directions.

\section{Related work}
\label{sec:relwork}

One of the pioneering works on VNF placement is \cite{RCohen15}, which~casts placement as a generalized assignment problem (GAP) and proposes a near-optimal solution based on bi-criteria approximation. Very recent works~\cite{feng2017approximation,agarwal2019vnf,poularakis2019joint} focus on the mutual influence of VNF placement and traffic routing. Others tackle the VNF placement problem through graph theory~\cite{ma2017traffic,draxler2018jasper} and set-covering~\cite{sang2017provably}, obtaining very good competitive ratios (constant in specific cases for~\cite{sang2017provably}).

In the context of MEC, some works tackle tasks different from sheer data processing; as an example, \cite{xu2018joint,chen2018edge}~aim at jointly optimizing computation and caching offloading between cloud-based and MEC-based infrastructure. Others focus on additional decisions that can be made in slicing scenarios, e.g., priority assignment in~\cite{malandrino2019getting}. A body of works considers incremental deployment, i.e., service requests arriving at different times: in this case, it is possible to share existing VNF instances~\cite{malandrino2019getting,lukovszki2016s,lukovszki2018approximate}, augment routing paths instead of computing them from scratch~\cite{lukovszki2016s,lukovszki2018approximate}, and minimize the difference between current and future network configuration~\cite{lukovszki2016s}.
Among the few works tackling non-functional requirements, \cite{zhao2018benders}~performs resilient VNF placement,  to achieve robustness to equipment failures. More recently, \cite{kamran2019deco}~considered the problem of jointly placing the VNFs and the data they need.

VNF placement, along with the closely-related problem of VNF chaining, has been studied in the software-defined networking and cloud-computing contexts as well. For instance, \cite{kulkarni2017nfvnice}~focuses on updating the  placement  in order to react to traffic changes, and \cite{sun2017nfp}~deals with the parallelization opportunities offered by VNF graphs. Other works \cite{guo2015shadow,poularakis2019joint} focus on the choice between MEC- and cloud-based computation resources, 
while \cite{cohen2019access}~studies which cache storage (i.e., MEC- or cloud-based) to access, balancing miss probability and cost. 

Several works aim at simplifying the problem of VNF placement by characterizing and/or predicting the traffic demand. In particular, \cite{bouet2018mobile}  exploits the spatial and temporal variability of traffic demand  to serve it with as little resources as possible; as for demand prediction, popular approaches include reinforcement learning~\cite{sciancalepore2018z}. In a similar spirit, \cite{han2019utility} estimates the resources needed by an incoming service request before deciding whether or not it shall be accepted.

Finally, a body of work addresses the slicing of the radio access network; in particular, \cite{melodia1}  proposes solutions that let different virtual operators use the radio resources without interfering,  while \cite{mancuso1}  develops a stochastic model to investigate the throughput and delay of a slice as functions of the cell parameters. Although such specific aspects are out of the scope of our work, we do tackle the problem of selecting radio technologies and points of access that honor the required KPI targets and minimize the cost. 

{\bf Novelty.} Remarkably, {\em none} of the existing works accounts for fundamental KPIs in network slicing such as  availability or reliability: due to their pioneering nature, such works  focus on traditional performance metrics, namely, service latency and/or network throughput.  
Although some  approaches could be extended to account for additional KPIs, such extensions would not be trivial and would, in general, jeopardize their complexity and/or competitive ratio properties. {\vRoute}, on the contrary, is designed from the start to support multiple, heterogeneous KPIs in an effective and efficient manner, beside accounting for all types of resources and their location, from the fog to the cloud. Our methodology combining graph theory and optimization is also unique, and provides an effective way to tradeoff optimality and complexity.

\section{System model}
\label{sec:model}

Our model concisely describes the two main components of mobile, slicing-based networks: the services they support (\Sec{sub-services}), and the computing and network resources they include (\Sec{sub-infrastructure}). Each of them is modeled through a graph -- the service graph and the physical graph, respectively. We then describe how such graphs can be combined in \Sec{sub-assign}.

\subsection{Services}
\label{sec:sub-services}

\begin{figure}
\centering
\includegraphics[width=1\columnwidth]{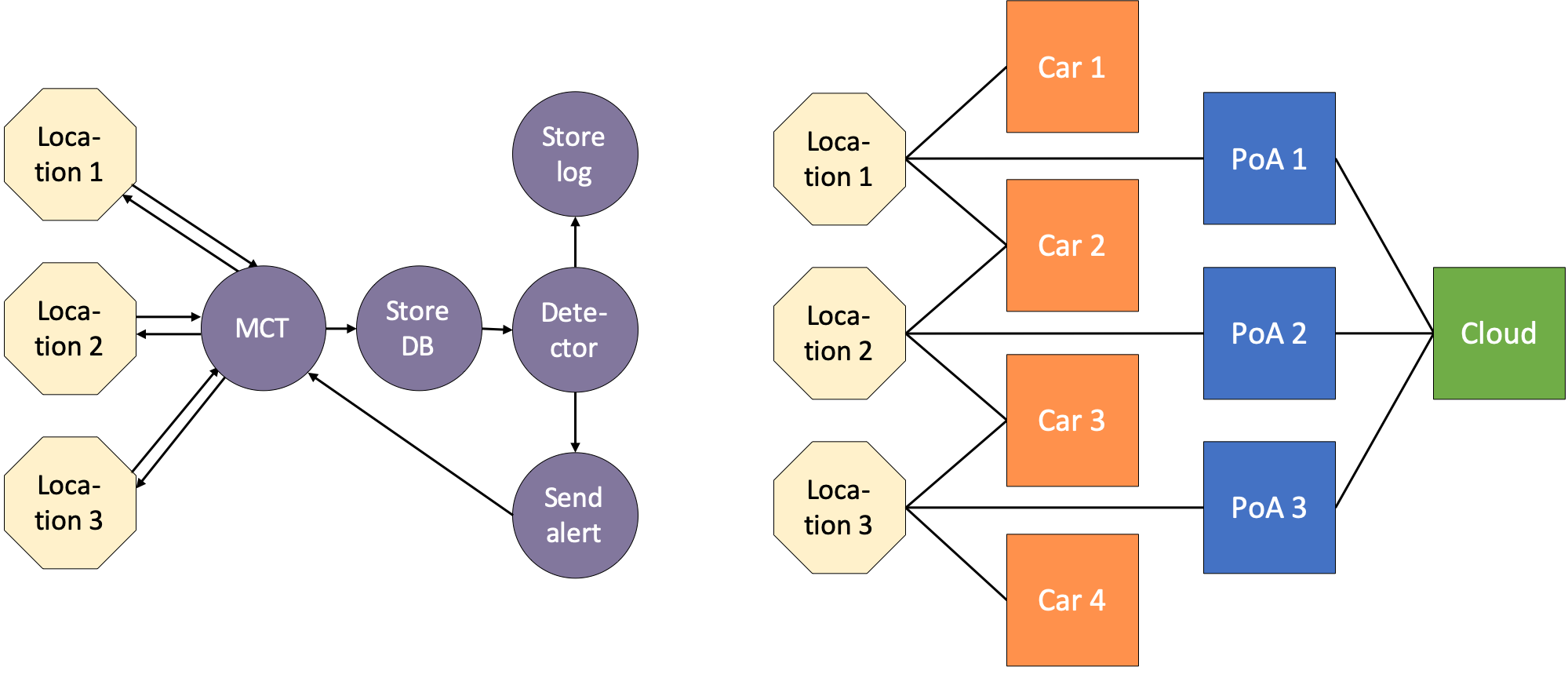}
\caption{\label{fig:graphs}Service (left) and physical (right) graphs corresponding to the example in \Fig{fresco}. Messages periodically sent by vehicles are collected through the Mobile Communication Transport (MCT), e.g., virtual eNB and vEPC, then  stored in a database and used for detecting vehicles on a collision course. The latter are warned by sending them an alert. In the service graph, vertices are endpoints  (yellow) and VNFs  (purple), edges are directed and correspond to flows $l$. In the physical graph, vertices are endpoints in $\Ec$ or  nodes in $\Cc$, and edges correspond to links in~$\Lc$; colors correspond to those in \Fig{fresco} and refer to the different resource locations: fog (orange), MEC (blue),  cloud (green).
} 
\end{figure}

A vertical service $s\in\Sc$  is described through a {\em service graph}  where vertices are VNFs,~$v\in\Vc$, and edges specify in which order the VNFs should process the related data traffic (i.e., how data shall be routed from a VNF instance running on a network node to the next). 
Note that VNFs can also represent database-related functionalities~\cite{kamran2019deco}, requiring storage resources: like other VNFs, they must be placed on a node and consume resources therein.

A service $s$ is associated with one or more {\em KPIs}, namely,
\begin{itemize}
  \item the required bandwidth, or expected traffic load $l$ to be  transferred and handled by the VNFs composing the  service;
    \item the maximum allowed delay $D(s)$;
    \item the minimum level of reliability $H(s)$;
   \item the required geographical availability at a subset of locations, $A(s)\subseteq\Ac$, where $\Ac=\{\alpha\}$~represents the set of all possible locations in the considered region. As an example, $A(s)$ can represent the  urban intersections where an automotive vertical wants to provide a safety service (\Fig{fresco}), or the  areas where robots should move within a warehouse. We refer to the combination of a service and a location as an {\em endpoint}~$e=(\alpha,s)\in\Ec\subseteq\Ac\times\Sc$;
    \item the lifetime (or temporal availability) $\varphi(e)\subseteq\Tc$, corresponding to a subset of all time steps~$\Tc$ during which the service must be available at endpoint $e$.
\end{itemize}
As foreseen by standards \cite{kpis}, services may be associated with one or more of these requirements, i.e., not all KPIs have to be specified for all services. Also, 
without loss of generality, we consider that the traffic associated with a service is generated at endpoint $e$ and has to be processed by the VNFs in the service graph; in \Fig{graphs}(left), this would correspond to uplink data transfers. Note however that our model is general and can also capture downlink as well as bidirectional traffic patterns. 

The quantity of traffic originated at endpoint~$e\in\Ec$, that has been processed last at VNF~$v_1$, and will be next processed at VNF~$v_2$ is denoted with $l(e,v_1,v_2)$ (with  $l(e,v,v)$ being the traffic that will be processed for the first time at $v$). 
After a traffic flow is processed at a VNF, the outgoing traffic can increase, decrease, or be split among several other VNFs, according to the service graph. Parameters~$\chi(v_1,v_2,v_3)$ express the fraction of the  traffic that was last processed (or originated) at $v_1\in\Vc\cup\Ec$, that is currently processed at $v_2$, and that will next be processed at $v_3$. For instance, if $v_2$ is a deep packet inspector, $\chi(v_1,v_2,v_3)=1$; but if $v_2$ is a firewall, then $\chi(v_1,v_2,v_3)\leq 1$.

The need for such~$\chi$-parameters is due to the fact that, as discussed in the previous example, there is no flow conservation on the service graph. Instead, the following {\em generalized flow conservation} law holds:
\begin{multline}
    l(e,v_2,v_3) = \sum_{\substack{v_1:\, v_1\ne v_2}} l(e,v_1,v_2) \chi(v_1,v_2,v_3)  \\
  +  l(e,v_2,v_2) \chi(e,v_2,v_3),\quad \forall v_2,v_3\in\Vc: v_2\ne v_3
    \label{eq:flow-conservation}
\end{multline}
The intuitive meaning of \Eq{flow-conservation} is that traffic traveling from VNF~$v_2$ to VNF~$v_3$ must either come from another VNF~$v_1$ and then it is transformed in~$v_2$ according to the $\chi$-coefficients (first term of the second member), or it has just originated at~$e$ and is processed for the first time at~$v_2$ (second term).

\subsection{Radio coverage and Fog/MEC/cloud resources}
\label{sec:sub-infrastructure}

Network nodes, with switching or computing capabilities, are denoted by $c \in \Cc$, while  endpoints, which are origins or destinations of service traffic, are denoted by $e \in \Ec$.  
 Nodes may be equipped with different resources, e.g., CPU or memory; the set of resources is identified by~$\Kc=\{\kappa\}$. The quantity of resource type~$\kappa$ available at node~$c$ is specified through parameters~$k(\kappa,c)$, hence, $k(\kappa,c)=0$ $\forall \kappa$ for pure network equipment like traditional, non-software, switches. Also, binary parameters~$R_i(c)$ express whether node~$c$ is equipped with radio interface~$i\in\Ic$ or not. A radio interface available at node $c$ determines which locations, hence endpoints, node $c$ covers -- an important feature of fog and MEC nodes. 

Radio coverage, fog, MEC, and cloud  resources can then be represented through a {\em physical graph} whose vertices are the network nodes and the endpoints, and the edges $(i,j)\in\Lc\subseteq(\Cc\cup\Ec)^2$ represent  the physical links connecting them, as per the network topology and the coverage provided by the radio interfaces.  
Each edge $(i,j)$ is associated with delay $D_{i,j}$ and traffic capacity $C_{i,j}$. 
Any node $c$ and link $(i,j)$ are associated with {\em reliability} values $\eta(c,t)$ and $\eta(i,j,t)$, respectively, which express the probability that a specific node or link works as intended at time~$t\in\Tc$. The fact that reliability values are time dependent models  real-world aspects like the fleeting quality of communication links involving fog nodes, e.g., robots or cars, as in \Fig{fresco}.

One of the main decisions to make through our model is where to process and route the service traffic. To this end, we introduce variables $\tau_{i,j}(e,v_1,v_2)$ representing  the flows over the physical graph, or, more specifically, the traffic originated at $e\in\Ec$, traversing $(i,j)\in\Lc$, last processed at  $v_1$, and to be next processed at  $v_2$. Such traffic can be either processed at~$j$, or just transiting through $j$; these two options are described by the two real variables~$p_{i,j}(e,v_1,v_2)$ and~$t_{i,j}(e,v_1,v_2)$ and by imposing:
    $\tau_{i,c}(e,v_1,v_2) = p_{i,c}(e,v_1,v_2) + t_{i,c}(e,v_1,v_2)$. 
Furthermore, the traffic going out of $c$ must be equal to the sum of that transiting through $c$ and  that just processed at~$c$:
\begin{multline}
    \sum_{(c,h)\in \Lc} \tau_{c,h} (e, v_2, v_3) \mathord{=} \hspace{-2mm} \sum_{(i,c) \in \Lc} \Big [ t_{i,c}(e, v_2, v_3) +p_{i,c}(e,v_2,v_2) \cdot\\
     \chi(e,v_2,v_3) + \sum_{v_1 \in \Vc} p_{i,c} (e, v_1, v_2) \chi(v_1, v_2, v_3) \Big ].
    \label{eq:conservation2}
\end{multline}
Finally, each physical link $(i,j)$ cannot carry more traffic than its capacity, i.e., $\sum_{e} \sum_{v_1, v_2} \tau_{i,j}(e, v_1, v_2) \le C_{i,j}$.

\subsection{Deploying VNFs and assigning  resources}
\label{sec:sub-assign}
A node can process traffic of a VNF if it hosts an instance of that VNF; this is modeled  through binary variables~$\rho(v,c)\in\{0,1\}$, expressing whether VNF~$v$ is deployed at node~$c$. Variables~$a_c(e,v,\kappa)$, instead, express the quantity of resources of type $\kappa$ that are assigned to the instance of VNF~$v$ deployed at node~$c$ and used for traffic generated at endpoint $e$. Such quantities cannot exceed the node capabilities, i.e., for any $c$ and $\kappa$,
 $\sum_{e \in \Ec} \sum_{v \in \Vc} a_c(e, v, \kappa) \leq k(\kappa,c)$. 

Importantly, for any $\kappa \in \Kc$, the quantity of traffic processed by $v$ at node~$c$ cannot exceed the ratio between the quantity~$a_c(e,v,\kappa)$ of resource type~$\kappa$ assigned to the VNF, and the quantity~$r_\kappa(v)$ of resource  type~$k$ needed by VNF~$v$ to process one unit of traffic:
\begin{equation}
\label{eq:resources}
\sum_{(i,c)\in\Lc}\sum_{e\in\Ec}\sum_{v_1\in\Vc}p_{i,c}(e,v_1,v_2) \leq \frac{a_c(e,v,\kappa)}{r_\kappa(v_2)} \quad \forall \kappa\in\Kc\,.
\end{equation}
Also,  node $c$'s resources can be assigned to a VNF $v$ only if the latter is deployed therein:
$a_c(e,v,\kappa)\leq \rho(v,c)k(\kappa,c)$, for any $c$, $\kappa$, and $v$.  
These conditions 
  imply that no traffic is processed at a node where no instance of a  VNF is deployed.

Last, we ensure that VNFs are placed only  at nodes where all the needed radio interface(s) are available, e.g., an MCT may work  only at nodes equipped with specific radio interfaces.  Thus, for any node $c$, interface $i$, and VNF $v$, we have:
    $\rho(v,c) r_i(v) \leq R_i(c)$,
where~$r_i(v)\in\{0,1\}$ are  parameters specifying whether interface~$i$ is needed by VNF $v$, and $R_i(c)$ specifies whether such an interface is available at $c$.

{\bf Matching service and physical flows.} 
Since our system model includes two graphs, we must ensure that service flows~$l$ and physical flows~$\tau$ match. To this end, we impose that the flow incoming the first VNF of a service graph corresponds to one or more traffic flows on the physical graph:
\begin{equation}
\label{eq:agree}
    l(e,v,v) = \sum_{(e,c) \in \Lc} \tau_{e,c} (e, v, v), \forall e\in \Ec, v\in \Vc.
\end{equation}
Once \Eq{agree} is met, then \Eq{flow-conservation} and 
\Eq{conservation2} ensure that the traffic on subsequent links is processed as specified by the $\chi$-parameters. 

\section{Problem formulation}
\label{sec:problem}
In this section, we formalize the problem of creating end-to-end network slices that meet all the required KPI targets (\Sec{sub-kpi}) while minimizing the total cost (\Sec{sub-obj}). The problem complexity is then discussed in \Sec{sub-complexity}. 

\subsection{Meeting service KPIs\label{sec:sub-kpi}}


To handle  more easily the service KPIs, we define a {\em string}, $w \in Wc$, over the physical graph as a sequence of physical links traversed by a flow, with the first  item of the string being an endpoint. Similarly to~\cite{qazi2013simple}, the possible strings can be pre-computed and stored for later usage.

Since a service flow can be split across different strings, we define $f(e,v_1,v_2,w)$ as the fraction of service flow $l(e,v_1,v_2)$ traversing  string $w$.  Clearly, such fractions must sum to 1.
We also introduce string-wise equivalents to $\tau_{i,j}(e,v_1,v_2)$, $t_{i,j}(e,v_1,v_2)$, and $p_{i,j}(e,v_1,v_2)$. Specifically,  $\tau_{i,j}(e,v_1,v_2,w)$ represents the traffic of service flow $l(e,v_1,v_2)$ traversing link $(i,j)$ on its journey through string $w\in\Wc$, and then impose 
  $  \tau_{i,j}(e,v_1,v_2) = \sum_{w\in\Wc} \tau_{i,j}(e,v_1,v_2,w)$. 
Similar definitions and conditions hold for $t_{i,j}(e,v_1,v_2,w)$ and $p_{i,j}(e,v_1,v_2,w)$.

Furthermore, the fraction of service flow over a certain string $w$ must match the physical traffic on the corresponding links, i.e., for all endpoints, VNFs $v_1$ and $v_2$, links, and strings, we have: 
  $  f(e,v_1,v_2,w) l(e,v_1,v_2) = \tau_{i,j}(e,v_1,v_2) \ind{w(i,j)}$, where 
$ \ind{w(i,j)}$ denotes that link $(i,j)\in\Lc$ 
belongs to $w$. 

\subsubsection{Service latency}
it has two components: network delay due to traffic traversing links and switches, and processing times at the nodes hosting VNF instances. Given endpoint~$e$, the average network delay  can be computed as the weighted sum of the delays associated with the individual strings taken by the traffic originated at $e$:
\begin{equation}
\label{eq:net-delay}
d_\text{net}(e)=\sum_{w\in\Wc}\sum_{v_1,v_2\in\Vc}f(e,v_1,v_2,w)\sum_{(i,j)\in w}D_{i,j}.
\end{equation}

As for the processing time, let~$\hat{f}_c(e,v_1,v_2)$ be the fraction of the service traffic  flow~$l(e,v_1,v_2)$  processed at the instance of VNF $v_2$ located at node~$c$.
Then the quantity of traffic~$\lambda_c(e,v_2)$ originating at~$e$ and processed at the instance of $v_2$ in $c$ is:  
\begin{equation}
\nonumber
\lambda_c(e,v_2)=\sum_{v_1\in\Vc}\hat{f}_c(e,v_1,v_2)l(e,v_1,v_2).
\end{equation}
Note that such traffic may come from different physical links.  

Next, we model VNF instances as M/M/1-PS queues (see, e.g., \cite{RCohen15}); the choice of the processor sharing (PS) policy closely emulates the behavior of a  multi-threaded application running on a virtual machine.  Hence,  the total processing time at the instance of~$v_2$ deployed at node~$c$ is:
$1/(a_c(e,v_2,\text{cpu}){-}r_\text{cpu}(v_2)\lambda_c(e,v_2))$.
Summing over all  flows, the total processing delay incurred by traffic originating at~$e$ can be written as:
\begin{equation}
\label{eq:proc-time}
d_{\text{proc}}(e){=}\hspace{-5mm}\sum_{v_1,v_2\in\Vc,c\in\Cc}\hspace{-5mm}\hat{f}_c(e,v_1,v_2)\frac{1}{a_c(e,v_2,\text{cpu})-r_\text{cpu}(v_2)\lambda_c(e,v_2)}.
\end{equation} 
Combining the above equation with \Eq{net-delay}, and recalling that $D(s)$ is the maximum target delay for service $s$, the service latency constraint for its endpoints can be stated as:
\begin{equation}
\label{eq:constr-latency}
d_\text{net}(e)+d_\text{proc}(e)\leq D(s),\quad\forall e\in\Ec.
\end{equation}

Finally, note that the relationship between assigned CPU and processing time in the expression of  $d_{\text{proc}}(e)$ also means that the CPU has a different role from the other types of resources. Indeed, for resources other than CPU, we can assign to each VNF instance exactly the amount needed to honor \Eq{resources}, as a greater amount would yield no benefit. With CPU, instead, there is an additional degree of freedom we can play with:  assigning more CPU results in shorter processing times, but higher costs.

\subsubsection{Service geographical availability}
by service availability requirements, all locations in~$A(s)\subseteq\Ac$ must be covered by service $s$. In other words, for all endpoints~$e=(\alpha,s)\colon\alpha\in A(s)$, there must be a link~$(e,c)$ on the physical graph to a node~$c$ that is equipped with a radio interface covering $\alpha$ and that runs (or it is connected to another node running) the first VNF of the service graph.  

\subsubsection{Service reliability and temporal availability} 
the reliability of a string can be computed as the product between the reliability values of all links and nodes belonging to it. We can therefore ensure that the reliability~$H(s)$ required for service $s$ is honored, by considering a weighted sum of the per-string reliability values. In symbols, $\forall e\in\Ec,t\in\varphi(e)$,
\begin{equation}
\prod_{v_1,v_2\in\Vc}\sum_{w\in\Wc}f(e,v_1,v_2,w)\prod_{(i,j)\in w}\eta(j,t)\eta(i,j,t)\geq H(s)\,.
\end{equation}
Note that imposing the above constraint for every time instant during the service lifetime also ensures that the target temporal availability  is met.

\subsection{Objective}
\label{sec:sub-obj}

As mentioned in \Sec{intro}, cost is one of the main concerns related to service virtualization and network slicing. Such cost mainly comes from using network and computation resources.  
To model this issue, we define:
\begin{itemize}
\item a fixed cost~$c_c(v)$, due to the creation at node~$c$ of a VNF instance $v$; this cost is null if an existing VNF instance can be reused; 
    \item a cost~$c_c(\kappa)$, incurred when using a unit resource~$\kappa$ at node~$c$;
    \item a cost~$c_{i,j}$, incurred when one  unit traffic traverses link~$(i,j)$.
\end{itemize}

Then, upon receiving a request to deploy a service instance $s$, we formulate the following cost-minimization problem:
\begin{multline}
\label{eq:obj}
\min
\sum_c\sum_{v} \left [ c_c(v)+\sum_{e}\sum_{\kappa}c_c(\kappa)a_c(e,v,\kappa) \right ] \\
+\sum_{(i,j)} \sum_{e} \sum_{v_1,v_2}c_{i,j}\tau_{i,j}(e,v_1,v_2) 
\end{multline}
subject to the constraints reported in Secs.\,\ref{sec:sub-services}--\ref{sec:sub-kpi}. 

We recall that  the endpoints $e$ to consider depend on the service and on its geographic availability requirements, while the VNFs are those specified by the service graph.
Furthermore, a solution to the above problem will always opt for reusing an existing instance of a VNF, whenever possible, as this would  nullify the instantiation cost $c_c(v)$.  

\subsection{Nature and complexity of the problem}
\label{sec:sub-complexity}

The problem of jointly making VNF placement and data routing decisions is notoriously hard; indeed, simpler versions thereof (considering one KPI only) have been proven to be NP-hard via reductions from the generalized assignment~\cite{RCohen15,feng2017approximation} and set covering~\cite{sang2017provably} problems. Thus, directly solving the above problem is impractical for all but very small instances.

We also observe that our problem can be seen as a more complex version of a multi-constrained path (MCP) problem, where the  cost (hence, the weight of the edges in the MCP graph) {\em changes} at every hop. Although known solutions to the MCP problem, e.g.,~\cite{xue2007finding}, are not applicable, such a similarity motivates us to propose an effective and efficient heuristic, called {\vRoute}, for which we  prove that:
\begin{itemize}
\item it provides high-quality VNF placement and data routing decisions, whose feasibility is guaranteed;
\item such decisions are made in polynomial time;
\item under mild homogeneity assumptions, decisions are optimal; 
\item in the general case, they can be arbitrarily close to the optimum. 
\end{itemize}

\section{The {\vRoute} solution}
\label{sec:algo}

Our solution includes three main steps. First, by leveraging the physical graph, we create a {\em decision graph}~$\widetilde{G}=(\widetilde{N},\widetilde{E})$, summarizing the  service deployment decisions that can be made and their effect on the KPIs (\Sec{sub-graph}). Then we translate this graph into an {\em expanded graph}, and use the latter, along with the service graph in \Sec{sub-services}, to identify a set of feasible decisions as well as to select, among them, the lowest-cost one (Secs.\,\ref{sec:sub-feasible} and \ref{sec:sub-cost}). 
For presentation clarity, we present {\vRoute} in the case where  the service graph is a chain starting from an endpoint~$e$ and including~$N$ VNFs~$v_1\dots v_N$, and only one  instance of each VNF can be placed. As discussed in \Sec{sub-gen}, both limitations can be dropped:  {\vRoute}  works with arbitrary service graphs requiring any number of instances of each VNF.

\subsection{The decision graph}
\label{sec:sub-graph}

Given the physical graph modeling the service endpoints and the fog, MEC, and cloud resources, we build the decision graph $\widetilde{G}$ with the aim to represent the possible service deployment decisions and  their effects on the service KPIs. 

As a preliminary step, we consider the {\em computation-capable} nodes in the physical graph  (hence, a subset of $\Cc$), and for each of them we create $(|\Vc|-1)$ replicas. 
Consistently, we create auxiliary  edges (i) connecting each node $c$ and its replicas in a chain fashion, and assign  them zero delay, infinite capacity, and reliability~$1$, and (ii) connecting  any replica of $c$ with any computing node $d$, for which a link $(c,d)\in \Lc$ exists.  
Crucially,  introducing  node replicas enables us to account for the possibility to deploy multiple VNFs at the same node without introducing self-loops in the decision graph. Indeed, as it will be more clear later, given that a VNF is placed in $c$, each replica thereof represents the possibility to deploy the next VNF again in $c$.

Let then $\widetilde{G}=(\widetilde{N},\widetilde{E})$ be the decision graph where: 
\begin{itemize} 
\item $\widetilde{N}$  includes the  endpoints in $\Ec$, and the  computation-capable nodes in the physical graph as well as their replicas;   
\item $\widetilde{E}$ is the set of (i) the aforementioned auxiliary links, and (ii) the virtual links (i.e., single physical links or sequences thereof) connecting the vertices in $\widetilde{N}$.  
\end{itemize} 
Every edge $(\tilde{n}_1,\tilde{n}_2)$ in  $\widetilde{E}$ representing a virtual link has the following properties:
\begin{itemize}
    \item its capacity~$\widetilde{C}_{\tilde{n}_1,\tilde{n}_2}$ is set to the minimum of the individual capacities of the physical links composing the virtual link;
    \item its delay~$\widetilde{D}_{\tilde{n}_1,\tilde{n}_2}$ is set to the sum of the individual delays of the physical  links composing the virtual link;
    \item its reliability~$\widetilde{\eta}_{\tilde{n}_1,\tilde{n}_2}$ is set to the product of the reliability values of physical links and nodes (both computation and pure-routing capable) included in the virtual link.
\end{itemize}

We stress that, when some services are already active in the network, we build the decision graph considering the {\em residual} capabilities of physical  links and nodes, i.e., those not assigned to already-running services. Similarly, in case of virtual links sharing the same physical links, their capacity is updated as traffic is allocated to the physical links.

\subsection{The expanded graph: finding decisions honoring availability and additive KPIs}
\label{sec:sub-feasible}

Given the decision graph~$\widetilde{G}$, our first purpose  is to identify a set of {\em feasible} service deployment decisions that are consistent with the target  KPIs. To this end, as a preliminary step, we ensure to meet the geographical and temporal availability requirements by pruning from  $\widetilde{G}$ the vertices and edges that do not satisfy such constraints. 

Then, for the additive KPIs,  
we proceed as follows.  
To any edge 
$(\tilde{n}_1,\tilde{n}_2)$ in the decision graph, we assign  a {\em multi-dimensional weight} $\tilde{w}(\tilde{n}_1,\tilde{n}_2)$, defined as:
\begin{equation}
\label{eq:weigth}
\tilde{w}(\tilde{n}_1,\tilde{n}_2)=\left(\frac{\widetilde{D}_{\tilde{n}_1,\tilde{n}_2}}{D(s)},\frac{\log \widetilde{\eta}_{\tilde{n}_1,\tilde{n}_2}}{\log H(s)}\right).
\end{equation}
The intuition behind  \Eq{weigth} is that the weight of edge~$(\tilde{n}_1,\tilde{n}_2)$ corresponds to the fraction of the target delay and reliability that will be ``consumed'' by taking that edge, i.e., by deploying a VNF at~$\tilde{n}_1$ and the subsequent one at~$\tilde{n}_2$. 
We stress that using logarithms in the second term of the weight allows us to translate a multiplicative performance index (namely, reliability)  into an additive one\footnote{It is easy to see that $\widetilde{\eta}_{\tilde{n}_1,\tilde{n}_2}  \widetilde{\eta}_{\tilde{n}_2,\tilde{n}_3} \geq H(s)$ translates into  $\frac{\log \widetilde{\eta}_{\tilde{n}_1,\tilde{n}_2} }{\log H(s)} + \frac{\log \widetilde{\eta}_{\tilde{n}_2,\tilde{n}_3} }{\log H(s)} \leq 1$.}. 

\begin{figure}
\centering
\includegraphics[width=1\columnwidth]{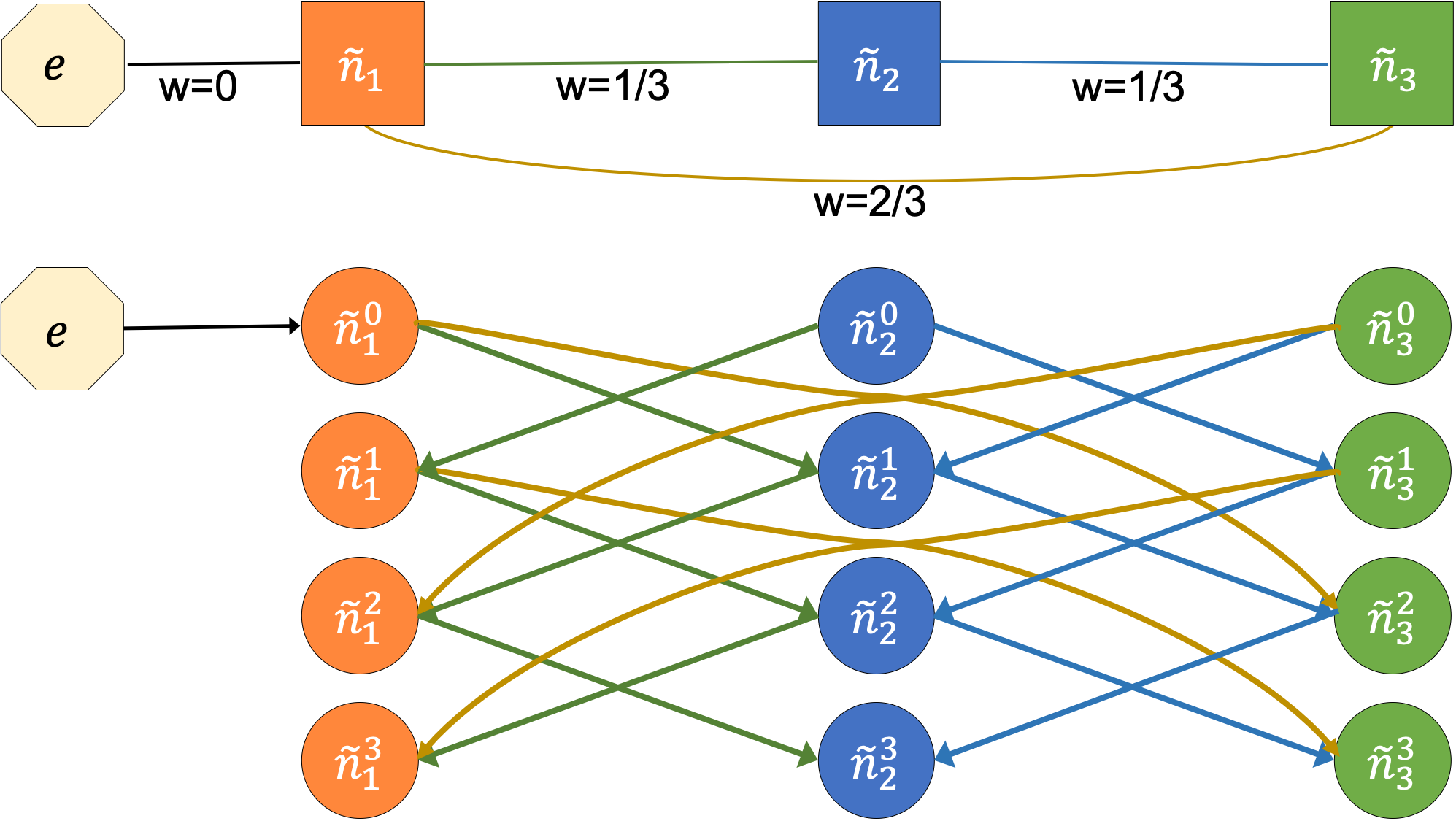}
\caption{Decision graph (top) and expanded graph (bottom) when only delay is considered as a KPI and $\gamma=3$. In the decision graph, edges~$(\tilde{n}_1,\tilde{n}_2)$ and~$(\tilde{n}_2,\tilde{n}_3)$ have  delay of 1~ms, while~$(\tilde{n}_1,\tilde{n}_3)$ has  delay of 2~ms (vertices representing replica nodes are omitted for simplicity); the target delay is 3~ms.
\label{fig:expanded}
} 
\end{figure}

Next, we take an approach inspired by~\cite{xue2007finding} and build a multi-dimensional, {\em expanded graph}. Specifically, given a positive integer value of resolution~$\gamma$:
\begin{enumerate}
\item for each vertex~$\tilde{n}$ in the decision graph, create $(\gamma+1)^2$ vertices\footnote{The exponent 2 corresponds to the number of additive KPIs.}  $\tilde{n}^{0,0}, \tilde{n}^{0,1}\ldots, \tilde{n}^{0,\gamma}\ldots,\tilde{n}^{\gamma,\gamma}$;
\item for every edge~$(\tilde{n}_1,\tilde{n}_2)\in\widetilde{E}$ with capacity~$\widetilde{C}_{\tilde{n}_1,\tilde{n}_2}$ greater or equal to the amount of traffic to process, create directed edges from each vertex~$\tilde{n}_1^{i,j}$ to vertex~$\tilde{n}_2^{i+\left\lceil\gamma w(\tilde{n}_1,\tilde{n}_2)[0]\right\rceil,j+\left\lceil\gamma w(\tilde{n}_1,\tilde{n}_2)[1]\right\rceil}$ (if such a vertex exists), where the two superscripts refer to delay and  reliability, respectively.
\end{enumerate}
We stress  that the expanded graph has {\em no weights} on its edges: the delay and reliability information that is expressed by weights in the decision graph is now represented by the topology of the expanded graph. A  one-dimensional (i.e., one-KPI) example of decision graph and corresponding expanded graph is depicted in \Fig{expanded}. 

Finally, we identify a set of possible service deployments, i.e., VNF-to-compute node assignments and the corresponding data routing.  
We do so by looking for the shortest paths\footnote{In our implementation, we use the Bellman-Ford algorithm.} in the expanded graph that (a) 
 begin at  endpoints, and (b) 
contain as many edges as there are VNFs to place. We underline that the latter is trivially required by the need to deploy all VNFs on the service graph, and by the fact that  placing more VNFs at a physical node is allowed thanks to the replica nodes and auxiliary edges, as discussed in \Sec{sub-infrastructure}. 

In the following, we make several fundamental remarks on the expanded graph and on the paths, hence,  the deployment decisions they correspond to. 
Given a KPI,  we  define as {\em depth} of a vertex in the expanded graph the quantity in the corresponding superscript; note that,  by construction (see point 1 above), the maximum value of depth is $\gamma$. Also, let the {\em steepness} of an edge be the difference in depth between its target and source vertices. Considering the one-KPI example in \Fig{expanded}(bottom), vertex~$\tilde{n}_1^0$ has depth~0, vertex~$\tilde{n}_3^2$ has depth~2, and the edge between the two has steepness~$2-0=2$, i.e., equal to $\lceil\gamma w(\tilde{n}_1,\tilde{n}_3)\rceil$. 

By construction, for a given KPI, the ratio between the steepness of an edge and~$\gamma$ is greater or equal to the weight component on the corresponding edge of the decision graph, which in turn is the fraction of the KPI target values consumed by making that decision (see \Eq{weigth}). As an example, considering edge~$(\tilde{n}_1^0,\tilde{n}_3^2)$ in \Fig{expanded}(bottom), we have:
\begin{equation}
\label{eq:steepness}
\frac{\text{steepness}}{\gamma}=\frac{2}{3}\geq w(\tilde{n}_1^0, \tilde{n}_3^2)=\frac{D_{\tilde{n}_1, \tilde{n}_3}}{D(s)}=\frac{2}{3}.
\end{equation} 
The observations above  allow us to state a very relevant property of the decisions corresponding to the paths on the expanded graph. 
\begin{lemma}
\label{lem:additive}
The decisions corresponding to any path on the expanded graph honor all additive KPIs.
\end{lemma}
\begin{IEEEproof}
By definition, the depth of a vertex corresponds to the total steepness of the path required to reach it from  endpoint~$e$. Given that  the maximum depth in the expanded graph is~$\gamma$,  there is {\em no path} with total steepness\footnote{The steepness of a path should be not confused with the length of a path.} greater than~$\gamma$. Thanks to the relation between weight and KPI targets (exemplified in \Eq{steepness}), this  implies that, given a path on the expanded graph, the sum of the weights of the corresponding edges in the decision graph cannot  exceed~$1$, i.e.,  
the corresponding decisions honor  additive KPIs (including, thanks to the logarithmic weights, reliability). 
\end{IEEEproof}

Importantly, the smaller the resolution~$\gamma$, the fewer the possible values of depth and steepness in the expanded graph, the fewer the levels of  consumption  of the KPI target values we are able to distinguish, which corresponds to introducing an error, akin to quantization. 
Indeed, $\gamma+1$ can be seen as the number of quantization levels\footnote{Using logarithms for reliability values, which are all typically very close to~1, is akin to performing adaptive quantization.} we admit: in the extreme case of~$\gamma=1$, all edges would have a steepness of~$1$, which also corresponds to exhausting the whole KPI target  in one hop. 
Such a quantization error may lead to discarding some feasible solutions, and thus, in the most general case, may jeopardize the optimality of {\vRoute}. However, two important facts stand out: {\em (i)} even enumerating all feasible paths in the decision graph is NP-hard, as proven in~\cite{xue2007finding}, hence, quantization is necessary; {\em (ii)} by increasing~$\gamma$, {\vRoute} can get {\em arbitrarily close to the optimum} (at the price of  higher complexity).

Finally, we remark that all paths on the expanded graph honor additive KPIs constraints,  {\em with the possible exception of delay}.  
Indeed, unlike other KPIs, whether or not the delay target is violated depends not only on the network latency, hence, the VNF placement, but also on the processing time, i.e., the quantity~$a_c(e,v,\text{cpu})$ of CPU assigned to each VNF, which in turn impacts the deployment cost. We can account for this important aspect thanks to the M/M/1-PS model used for \Eq{proc-time}. 
In particular, below we show how to determine, given a possible deployment, whether there is a CPU assignment consistent with the target delay, and the cost thereof. 

\subsection{Minimizing the cost}
\label{sec:sub-cost}

We now need (i) for every path found in \Sec{sub-feasible}, to identify the minimum-cost CPU assignment, i.e., the optimal values of the  $a_c(e,v,\text{cpu})$~variables -- if such an assignment exists --, and (ii) to determine the path that minimizes the overall cost.

To this end, for each  path (hence,  for a fixed $e$ and for VNFs~$v_1\dots v_N$ to be deployed at computing nodes~$\tilde{n}_1\dots\tilde{n}_N$, respectively), 
we solve the following  problem:
\begin{eqnarray}
\label{eq:mini-obj}
&& \min\sum_{\tilde{n},v}a_{\tilde{n}}(e,v,\text{cpu})c_{\tilde{n}}(\text{cpu}) \quad \quad \mbox{s.t.}\\
&& \hspace{-0.5cm} \sum_{\tilde{n}_1,\tilde{n}_2}\hspace{-1mm}\left(D_{\tilde{n}_1,\tilde{n}_2}\mathord{+}\frac{1}{a_{\tilde{n}_2}(e,v_2,\text{cpu}) \mathord{-}r_{\text{cpu}}(v_2)\lambda_{\tilde{n}_2}(e,v_2)}\right) \mathord{\leq} D(s). \nonumber
\end{eqnarray}
If the problem above is infeasible for a given path, then that path (and the corresponding decisions) is incompatible with the delay target KPI and must be discarded.

Once the problem in \Eq{mini-obj} is solved for all paths  identified  in the expanded graph, we compute the total cost associated with each path (including all components defined in \Sec{sub-obj})  and select and enact the lowest-cost deployment, thus fulfilling {\vRoute}'s purpose.  
Importantly, the problem is {\em convex}, hence, it can be efficiently solved in polynomial time~\cite{boyd}. The proof simply follows from observing that the objective in \Eq{mini-obj} is linear and the second derivatives of the delay constraint 
are positive in the decision variables, hence, the constraint itself is convex.

\subsection{General scenarios}
\label{sec:sub-gen}

We now show how {\vRoute} tackles arbitrary scenarios.

\subsubsection{Arbitrary service graphs}

If the service graph is more complex than a chain, as in \Fig{graphs}(left), we can proceed by (i) decomposing the graph into a set of chains (e.g., one in uplink, from the MCT to the DB, and one in downlink, from the detector back to the MCT). {\vRoute} is then applied subsequently to each chain, and the deployment decisions are cascaded. The case where multiple endpoints have to be covered, as in \Fig{graphs}(left), is handled in the same way.

\subsubsection{Multiple VNF instances}

If the problem described in \Sec{sub-cost} is infeasible for {\em all} possible paths found in \Sec{sub-feasible}, a possible reason could be the need to split the processing burden across  multiple instances of the same VNF. This case is handled by first  identifying the bottleneck VNF, i.e., taking the longest to process the service traffic, and then increasing by one the number of instances of that VNF in the service graph.  
{\vRoute} is then re-run on the modified service graph.

\section{{\vRoute} analysis}
\label{sec:analysis}

In this section, we prove several properties about {\vRoute}; 
we start with the most essential  aspect related to its effectiveness, i.e., its ability to meet all service KPIs.
\begin{property}
\label{prop:correct}
{\vRoute}'s decisions honor all KPI targets.
\end{property}
\begin{IEEEproof}
By \Lem{additive}, all decisions honor the additive KPIs. 
Concerning delay, it is guaranteed that such a KPI target is met, thanks to the delay constraint  imposed while  performing the CPU assignment. As noted in \Sec{sub-cost}, decisions resulting in an infeasible problem are discarded, hence, the selected decision   honors the  delay target. 
Finally, the availability constraints are satisfied through the initial selection of the vertices of the decision graph (see \Sec{sub-feasible}). 
\end{IEEEproof}

Next, we address the computational complexity of {\vRoute}. 
\begin{property}
\label{prop:poly}
The worst-case computational complexity of {\vRoute} is polynomial.
\end{property}
\begin{IEEEproof}
To prove the property, we show that each of the steps described in \Sec{algo} has a polynomial runtime. Specifically, 
(i) creating the expanded graph (\Sec{sub-graph}) requires creating at most~$\gamma^2(|\Vc||\Cc|+|\Ec|)$ nodes and 
 at most~$\gamma^2 |\Vc||\Lc|$ edges, where $|\Vc|$ is the number of VNFs specifying the service and, given the service,  is a constant. 
(ii)  Finding the possible decisions (\Sec{sub-feasible}) implies computing the shortest paths between any endpoint (i.e., vertex meeting the availability constraints) and any other  node in the expanded graph, which, in the worst case, has complexity \cite{seidel1995all} $o(n^{2.3})$ with $n$ being  the number of nodes in the expanded graph. 
(iii) Computing the optimal CPU assignments (\Sec{sub-cost}) has cubic complexity~\cite{boyd} in the problem size; indeed, convex problems are routinely solved in embedded computing scenarios.
\end{IEEEproof} 

Finally, under reasonable assumptions about the homogeneity of the physical graph, we can prove that {\vRoute} can actually return the optimal solution. 
\begin{property}
\label{prop:opti}
If all links and nodes have the same capabilities and cost, then the output of {\vRoute} is optimal.
\end{property}
\begin{IEEEproof}
There is only one point in the procedure we described where, in general scenarios, we may overlook the optimal solution. 
As remarked in \Sec{sub-feasible},  finite $\gamma$ values may cause a quantization-like error: in general scenarios, only for $\gamma \to \infty$, we could consider all possible ways to move from one node of the decision graph to another.  In the special case of homogeneous links and nodes, however, no such different possibilities exist, hence, taking a finite value of~$\gamma$ is enough to consider all possible choices the system offers and to make an optimal decision. Note that restricting our attention to shortest paths on the expanded graph does not harm optimality, as adding hops implies consuming a higher (or equal at best) fraction of KPI targets and cannot decrease the cost. 
\end{IEEEproof}

\section{Numerical results}
\label{sec:results}

Here, we first  focus on a small-scale scenario and a robot-based  smart factory application (\Sec{sub-small}), and compare the performance of {\vRoute} against the optimum obtained via brute force. 
Then  we move to a large-scale scenario and  real-world automotive application (\Sec{sub-large}), and we characterize how the quantity of traffic to serve and the maximum delay impact the decisions made by {\vRoute}.

\subsection{Small-scale scenario: comparison against the optimum}
\label{sec:sub-small}

\begin{figure}
\centering
\includegraphics[width=.8\columnwidth]{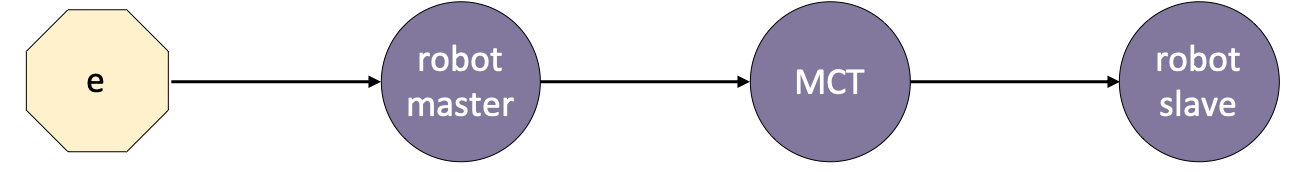}
\caption{Service graph specifying the robot-based, smart factory application.
\label{fig:robots}
} 
\end{figure}

\begin{figure*}
\centering
\includegraphics[width=.3\textwidth]{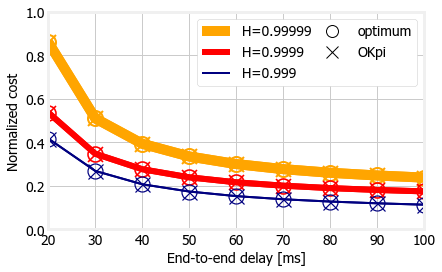}
\includegraphics[width=.3\textwidth]{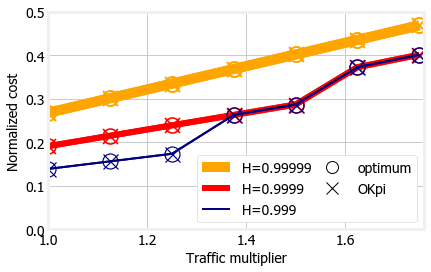}
\includegraphics[width=.3\textwidth]{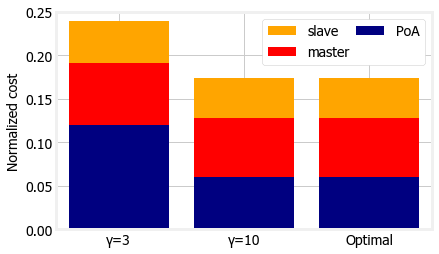}
\caption{Small-scale scenario: cost as a function of the maximum delay (left) and of the traffic load (center), for different values of target reliability; cost breakdown (right) when the target reliability is~0.999, the maximum delay is 50~ms, the traffic multiplier is~1, and $\gamma$~varies.
    \label{fig:results-a}
}
\centering
\includegraphics[width=.3\textwidth]{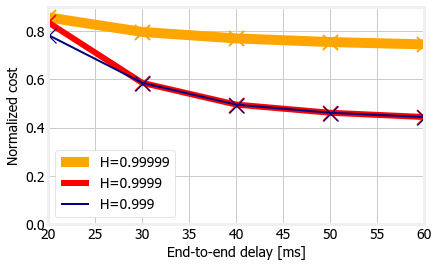}
\includegraphics[width=.3\textwidth]{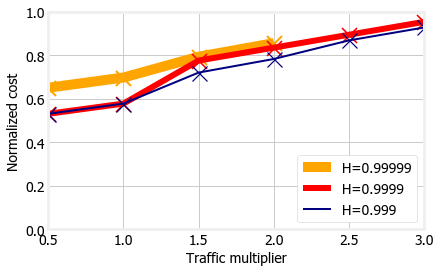}
\includegraphics[width=.3\textwidth]{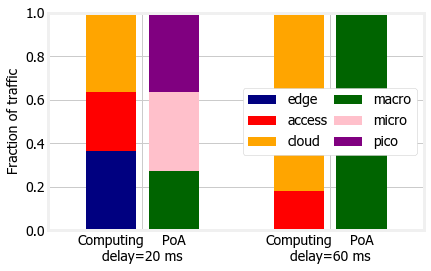}
\caption{Large-scale scenario: cost as a function of the maximum delay (left) and of the traffic load (center), for different values of target reliability; fraction of traffic (right) traversing different PoAs and computing nodes when the target reliability is~0.999, the traffic multiplier is~1, and the target delay varies.
    \label{fig:results-b}
}
\end{figure*}

We consider the robot-based, smart factory application~\cite{coral-robots}, whose service graph is depicted in \Fig{robots}. A room (hence, an endpoint) contains three robots, with different levels of reliability: $\eta(\text{robo1})=0.999999$, $\eta(\text{robo2})=0.99999$, and $\eta(\text{robo3})=0.9999$. Two of these three robots must be used to perform an operation, hence, run the {robo-master} and {robo-slave} VNFs. The communication between the two selected robots can take place through three types of PoAs, with different levels of reliability 
(micro-cell: 0.999999, 
pico-cell:  0.99999, 
femto-cell:  0.9994), and costs as reported in \cite{costs}. The offered traffic is 1~Mb/s per robot, as specified in~\cite{coral-robots}.

\Fig{results-a} depicts the results when {\vRoute}'s resolution is set to~$\gamma=10$. A first aspect we are interested in is the relationship between the target KPIs and cost: as we can see from \Fig{results-a}(left) and \Fig{results-a}(center), a longer allowable delay results in a lower cost; conversely, a higher traffic load or a higher target reliability both result in higher costs. Intuitively, this is due to the fact that cheaper resources (e.g., robot~3) tend to have lower reliability and/or capacity, hence, it is impossible to use them when the KPI targets become very strict.

Interestingly, in both \Fig{results-a}(left) and \Fig{results-a}(center), {\vRoute} matches the optimum in {\em all} cases. Indeed, as discussed in \Sec{sub-feasible}, {\vRoute}  always matches the optimum if the resolution~$\gamma$ is high enough; in the small-scale scenario we consider for \Fig{results-a},  $\gamma=10$ is sufficient to this end.

\Fig{results-a}(right) shows the effect of setting a lower resolution, namely,~$\gamma=3$. As we can see by comparing the left and center bars, a lower value of~$\gamma$ results in suboptimal, higher-cost decisions. Specifically, the difference is due to the fact that, when~$\gamma=3$, a higher-cost PoA is selected, namely, the pico-cell {\em in lieu} of the femto-cell. 
This happens because, for $\gamma=3$, the edges corresponding to the femto-cell in the expanded graph have steepness~$\left\lceil\gamma\frac{\log 0.9994}{\log{0.999}}\right\rceil=2$. Considering that (i) all other edges have steepness~1 and (ii)  {\vRoute}  seeks for paths composed of three edges (same as the number of VNFs to place) with a total steepness not exceeding~$\gamma=3$,  the  edges corresponding to the femto-cell will never be selected, hence, the corresponding decision is never considered. 
In summary, as discussed in the previous sections, using a too-low $\gamma$ made us overlook a feasible -- and, in this case, optimal -- solution.

\begin{figure}
\centering
\includegraphics[width=0.98\columnwidth]{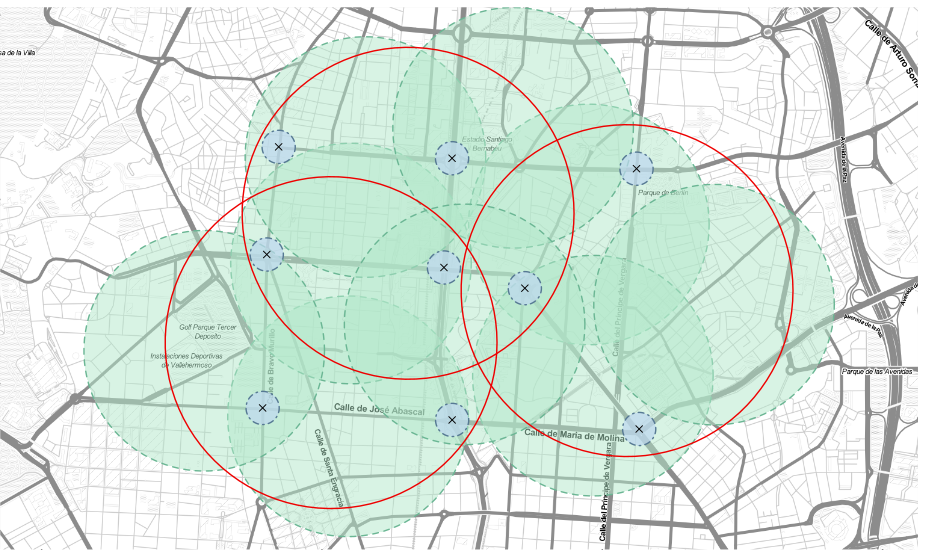}
\caption{Road topology used in the large-scale scenario. The nine crossings correspond to endpoints; red, green, and blue circles represent the coverage areas of macro-, micro- and pico-cells, respectively.
\label{fig:crossings}
} 
\end{figure}

\subsection{Large-scale scenario: impact of traffic and delay}
\label{sec:sub-large}

Our large-scale scenario depicts a urban environment where the vehicle collision avoidance system depicted in \Fig{fresco}~\cite{ica_CAMBIAPERCAMERAREADY} has to be provided. Based on a real-world road topology (see \Fig{crossings}), a total of 9 intersections (hence, endpoints) are covered by a combination of PoAs, namely, macro-, micro- and pico-cells, whose coverage is shown in \Fig{crossings}. 

Different PoAs have different reliability, latency, and cost values, as reported in \Tab{vehicular}. The front- and back-haul network topology is based on~\cite{martin2019modeling} and ITU standard~\cite{itu-topo}, and includes MEC, aggregation and core nodes with features as  summarized in \Tab{vehicular} \cite{d14-5gt}. The service graph to deploy is represented in \Fig{graphs}(left), the total service traffic is 1.5\,Mb/s, and   $\gamma$ is set to 40 (higher values do not improve the performance).

\begin{table}[tbh]
\caption{
Large-scale scenario: points of access and computing nodes
\label{tab:vehicular}
} 
\scriptsize
\begin{tabularx}{\columnwidth}{|X|r|r|r|r|}
\hline
{\bf Item} & {\bf Reliability} & {\bf Latency} & {\bf Cost}\\
\hline\hline
\multicolumn{4}{|>{\hsize=\columnwidth}c|}{Points of access}\\
\hline
macro-cell & 0.99999999 & 6~ms & 1.02~USD/Gbit  \\
\hline
micro-cell & 0.9999999 & 3~ms & 2.31~USD/Gbit\\
\hline
pico-cell & 0.999999 & 2~ms & 3.80~USD/Gbit\\
\hline
\multicolumn{4}{|>{\hsize=\columnwidth}c|}{Computing nodes}\\
\hline
cloud ring (Azure DataBox) & 0.99999999 & 8~ms & 2.23~USD/Gbit\\
\hline
aggregation ring (PowerEdge) & 0.9999999 & 3~ms & 5.23~USD/Gbit\\
\hline
MEC ring (small data center) & 0.999999 & 1~ms & 10.47~USD/Gbit  \\
\hline
\end{tabularx}
\end{table}

\Fig{results-b}(left) shows that, as one might expect, a shorter target delay results in higher costs. It is also interesting to observe the behavior of the intermediate curve, corresponding to~$H(s)=0.9999$: when the target delay is very short, its associated cost is almost the same as for $H(s)=0.999999$ case; as the target delay increases, its cost drops to the same level as the~$H(s)=0.999$ case. This bespeaks the complexity of the decisions {\vRoute} has to make, and their sometimes counter-intuitive effects.

In \Fig{results-b}(center), the traffic load is multiplied by a factor ranging between~0.5 and~3. We can again observe that to a higher traffic corresponds a higher cost, even though the growth is less than linear, owing to the fixed costs described in \Sec{sub-obj}. Also notice how the yellow curve in \Fig{results-b}(center), corresponding to the highest reliability level, stops at a multiplier of~2: for higher traffic demands, the network capacity is insufficient to provide the service with the required reliability.

\Fig{results-b}(right) shows which PoAs and computing nodes are selected for the minimum and maximum target delay values. Interestingly, in the presence of tight delay constraints, different PoAs and resources  are all used (left bars). On the contrary, for the largest target delay, the cheapest options -- cloud and macro-cells -- are preferred.

\section{Conclusion and future work}
\label{sec:conclusion}
We  identified in the support for a limited set of KPIs one of the main shortcomings of present-day approaches to network slicing. To address this issue, we proposed {\vRoute}, an efficient and effective solution strategy able to jointly make PoA selection, VNF placement, and data routing decisions, while natively accounting for all KPIs and for all resources, from the fog to the cloud. 
Importantly,  {\vRoute} draws on a novel methodology that blends together graph theory and optimization in a unique manner, and exhibits several desirable properties. Among these, we showed that   
{\vRoute} has polynomial computational complexity and  its performance can get arbitrarily close to the optimum. Our performance evaluation, carried out using two real-world scenarios, confirms that {\vRoute} closely matches the optimum and consistently provides very good performance.

Future work will focus on enhancing the expanded graph presented in \Sec{sub-feasible} by allowing edge steepness to take arbitrary real values, and on characterizing 
the competitive ratio of {\vRoute} as a function of the parameter $\gamma$.


\clearpage

\bibliographystyle{IEEEtran}
\bibliography{refs}

\begin{thebibliography}{10}
\providecommand{\url}[1]{#1}
\csname url@samestyle\endcsname
\providecommand{\newblock}{\relax}
\providecommand{\bibinfo}[2]{#2}
\providecommand{\BIBentrySTDinterwordspacing}{\spaceskip=0pt\relax}
\providecommand{\BIBentryALTinterwordstretchfactor}{4}
\providecommand{\BIBentryALTinterwordspacing}{\spaceskip=\fontdimen2\font plus
\BIBentryALTinterwordstretchfactor\fontdimen3\font minus
  \fontdimen4\font\relax}
\providecommand{\BIBforeignlanguage}[2]{{%
\expandafter\ifx\csname l@#1\endcsname\relax
\typeout{** WARNING: IEEEtran.bst: No hyphenation pattern has been}%
\typeout{** loaded for the language `#1'. Using the pattern for}%
\typeout{** the default language instead.}%
\else
\language=\csname l@#1\endcsname
\fi
#2}}
\providecommand{\BIBdecl}{\relax}
\BIBdecl

\bibitem{etsi-mec-wp2}
{ETSI}. (2018) {MEC in 5G networks}.
  \url{https://www.etsi.org/images/files/ETSIWhitePapers/etsi_wp28_mec_in_5G_FINAL.pdf}.

\bibitem{zhao2018benders}
P.~Zhao and G.~D{\'a}n, ``{A Benders decomposition approach for resilient
  placement of virtual process control functions in mobile edge clouds},''
  \emph{IEEE Transactions on Network and Service Management}, 2018.

\bibitem{malandrino2019getting}
F.~Malandrino and C.-F. Chiasserini, ``{Getting the Most Out of Your VNFs:
  Flexible Assignment of Service Priorities in 5G},'' in \emph{IEEE WoWMoM},
  2019.

\bibitem{sang2017provably}
Y.~Sang, B.~Ji, G.~R. Gupta, X.~Du, and L.~Ye, ``Provably efficient algorithms
  for joint placement and allocation of virtual network functions,'' in
  \emph{IEEE INFOCOM}, 2017.

\bibitem{pyramid}
``{IMT Vision: Framework and overall objectives of the future development of
  IMT for 2020 and beyond},'' \emph{ITU Recommendation M.2083-0}, 2015.

\bibitem{ica_CAMBIAPERCAMERAREADY}
Q.-H. Nguyen, M.~Morold, K.~David, and F.~Dressler, ``{Adaptive Safety Context
  Information for Vulnerable Road Users with MEC Support},'' in \emph{IEEE/IFIP
  WONS}, 2019.

\bibitem{lukovszki2018approximate}
T.~Lukovszki, M.~Rost, and S.~Schmid, ``Approximate and incremental network
  function placement,'' \emph{Elsevier Journal of Parallel and Distributed
  Computing}, 2018.

\bibitem{cohen2019access}
I.~Cohen, G.~Einziger, R.~Friedman, and G.~Scalosub, ``{Access Strategies for
  Network Caching},'' in \emph{IEEE INFOCOM}, 2019.

\bibitem{draxler2018jasper}
S.~Dr{\"a}xler, H.~Karl, and Z.~{\'A}. Mann, ``{Jasper: Joint optimization of
  scaling, placement, and routing of virtual network services},'' \emph{IEEE
  Transactions on Network and Service Management}, 2018.

\bibitem{poularakis2019joint}
K.~Poularakis, J.~Llorca, A.~M. Tulino, I.~Taylor, and L.~Tassiulas, ``{Joint
  Service Placement and Request Routing in Multi-cell Mobile Edge Computing
  Networks},'' in \emph{IEEE INFOCOM}, 2019.

\bibitem{xu2018joint}
J.~Xu, L.~Chen, and P.~Zhou, ``Joint service caching and task offloading for
  mobile edge computing in dense networks,'' in \emph{IEEE INFOCOM}, 2018.

\bibitem{agarwal2019vnf}
S.~Agarwal, F.~Malandrino, C.~F. Chiasserini, and S.~De, ``{VNF Placement and
  Resource Allocation for the Support of Vertical Services in 5G Networks},''
  \emph{IEEE/ACM Transactions on Networking}, 2019.

\bibitem{RCohen15}
R.~Cohen, L.~Lewin-Eytan, J.~S. Naor, and D.~Raz, ``Near optimal placement of
  virtual network functions,'' in \emph{IEEE INFOCOM}, 2015.

\bibitem{feng2017approximation}
H.~Feng, J.~Llorca, A.~M. Tulino, D.~Raz, and A.~F. Molisch, ``{Approximation
  algorithms for the NFV service distribution problem},'' in \emph{IEEE
  INFOCOM}, 2017.

\bibitem{ma2017traffic}
W.~Ma, O.~Sandoval, J.~Beltran, D.~Pan, and N.~Pissinou, ``{Traffic aware
  placement of interdependent NFV middleboxes},'' in \emph{IEEE INFOCOM}, 2017.

\bibitem{chen2018edge}
M.~Chen, Y.~Hao, L.~Hu, M.~S. Hossain, and A.~Ghoneim, ``{Edge-CoCaCo: Toward
  joint optimization of computation, caching, and communication on edge
  cloud},'' \emph{IEEE Wireless Communications}, 2018.

\bibitem{lukovszki2016s}
T.~Lukovszki, M.~Rost, and S.~Schmid, ``It's a match!: Near-optimal and
  incremental middlebox deployment,'' \emph{ACM SIGCOMM Computer Communication
  Review}, 2016.

\bibitem{kamran2019deco}
K.~Kamran, E.~Yeh, and Q.~Ma, ``Deco: Joint computation, caching and forwarding
  in data-centric computing networks,'' in \emph{ACM MobiHoc}, 2019.

\bibitem{kulkarni2017nfvnice}
S.~G. Kulkarni, W.~Zhang, J.~Hwang, S.~Rajagopalan, K.~Ramakrishnan, T.~Wood,
  M.~Arumaithurai, and X.~Fu, ``{NFVnice: Dynamic backpressure and scheduling
  for NFV service chains},'' in \emph{ACM SIGCOMM}, 2017.

\bibitem{sun2017nfp}
C.~Sun, J.~Bi, Z.~Zheng, H.~Yu, and H.~Hu, ``{NFP: Enabling network function
  parallelism in NFV},'' in \emph{ACM SIGCOMM}, 2017.

\bibitem{guo2015shadow}
Y.~Guo, A.~L. Stolyar, and A.~Walid, ``Shadow-routing based dynamic algorithms
  for virtual machine placement in a network cloud,'' \emph{IEEE Transactions
  on Cloud Computing}, 2015.

\bibitem{bouet2018mobile}
M.~Bouet and V.~Conan, ``Mobile edge computing resources optimization: a
  geo-clustering approach,'' \emph{IEEE Transactions on Network and Service
  Management}, 2018.

\bibitem{sciancalepore2018z}
V.~Sciancalepore, F.~Z. Yousaf, and X.~Costa-Perez, ``{z-TORCH: An automated
  NFV orchestration and monitoring solution},'' \emph{IEEE Transactions on
  Network and Service Management}, 2018.

\bibitem{han2019utility}
B.~Han, V.~Sciancalepore, D.~Feng, X.~Costa-Perez, and H.~D. Schotten, ``{A
  Utility-Driven Multi-Queue Admission Control Solution for Network Slicing},''
  in \emph{IEEE INFOCOM}, 2019.

\bibitem{melodia1}
S.~{D'Oro}, F.~{Restuccia}, A.~{Talamonti}, and T.~{Melodia}, ``The slice is
  served: Enforcing radio access network slicing in virtualized {5G} systems,''
  in \emph{IEEE INFOCOM}, 2019.

\bibitem{mancuso1}
V.~{Mancuso}, P.~{Castagno}, M.~{Sereno}, and M.~A. {Marsan}, ``Slicing cell
  resources: The case of {HTC and MTC} coexistence,'' in \emph{IEEE INFOCOM},
  2019.

\bibitem{kpis}
{5G PPP}. {Key Performance Indicators}. \url{https://5g-ppp.eu/kpis/}.

\bibitem{qazi2013simple}
Z.~A. Qazi, C.-C. Tu, L.~Chiang, R.~Miao, V.~Sekar, and M.~Yu, ``{SIMPLE-fying
  middlebox policy enforcement using SDN},'' \emph{ACM SIGCOMM computer
  communication review}.

\bibitem{xue2007finding}
G.~Xue, A.~Sen, W.~Zhang, J.~Tang, and K.~Thulasiraman, ``{Finding a path
  subject to many additive QoS constraints},'' \emph{IEEE/ACM Transactions on
  Networking}, 2007.

\bibitem{boyd}
S.~Boyd and L.~Vandenberghe, \emph{Convex optimization}.\hskip 1em plus 0.5em
  minus 0.4em\relax Cambridge university press, 2004.

\bibitem{seidel1995all}
R.~Seidel, ``On the all-pairs-shortest-path problem in unweighted undirected
  graphs,'' \emph{Journal of computer and system sciences}, 1995.

\bibitem{coral-robots}
``{5G-CORAL initial system design, use cases, and requirements},''
  \emph{5G-CORAL deliverable 1.1}, 2018,
  \url{http://5g-coral.eu/wp-content/uploads/2018/04/D1.1_final7760.pdf}.

\bibitem{costs}
V.~Nikolikja and T.~Janevski, ``A cost modeling of high-capacity {LTE-A}dvanced
  and {IEEE 802.11ac} based heterogeneous networks, deployed in the 700 {MHz,
  2.6 GHz and 5 GHz} bands,'' \emph{Procedia Computer Science}, 2014.

\bibitem{martin2019modeling}
J.~Mart{\'\i}n-P{\'e}rez, L.~Cominardi, C.~J. Bernardos, A.~De~la Oliva, and
  A.~Azcorra, ``Modeling mobile edge computing deployments for low latency
  multimedia services,'' \emph{IEEE Transactions on Broadcasting}, 2019.

\bibitem{itu-topo}
``{Consideration on 5G transport network reference architecture and bandwidth
  requirements},'' \emph{ITU Contribution 0462}, 2018.

\bibitem{d14-5gt}
``{Final system design and techno-economic analysis},'' \emph{5G-TRANSFORMER
  deliverable D1.4}, 2019.

\end{thebibliography}

\end{document}